# Reciprocal polarization imaging of complex media

Zhineng Xie[1], Guowu Huang[1], Weihao Lin[1], Xin Jin[1], Yifan Ge[1], Yansen Hu[1], Xiafei Qian[2] and Min Xu[1,3,*]

[1]Institute of Lasers and Biomedical Photonics, Biomedical Engineering College, Wenzhou Medical University, Wenzhou, Zhejiang, 325035, China.
[2]Chengbei District, Hangzhou First People's Hospital, Hangzhou, Zhejiang, 310000, China.
[3]Department of Physics and Astronomy, Hunter College and the Graduate Center, The City University of New York, 695 Park Avenue, New York, NY 10065, USA.
* minxu@hunter.cuny.edu

**Abstract:** The vectorial evolution of polarized light interaction with a medium can reveal its microstructure and anisotropy beyond what can be obtained from scalar light interaction. Anisotropic properties (diattenuation, retardance, and depolarization) of a complex medium can be quantified by polarization imaging by measuring the Mueller matrix. However, polarization imaging in the reflection geometry, ubiquitous and often preferred in diverse applications, has suffered a poor recovery of the medium's anisotropic properties due to the lack of suitable decomposition of the Mueller matrices measured inside a backward geometry. Here, we present reciprocal polarization imaging of complex media after introducing reciprocal polar decomposition for backscattering Mueller matrices. Based on the reciprocity of the optical wave in its forward and backward scattering paths, the anisotropic diattenuation, retardance, and depolarization of a complex medium are determined by measuring the backscattering Mueller matrix. We demonstrate reciprocal polarization imaging in various applications for quantifying complex non-chiral and chiral media (birefringence resolution target, tissue sections, and glucose suspension), uncovering their anisotropic microstructures with remarkable clarity and accuracy. We also highlight types of complex media that Lu-Chipman and differential decompositions of backscattering Mueller matrices lead to erroneous medium polarization properties, whereas reciprocal polar decomposition recovers properly. Reciprocal polarization imaging will be instrumental in imaging complex media from remote sensing to biomedicine and will open new applications of polarization optics in reflection geometry.

## 1 Introduction

The interaction of light with a medium provides a non-invasive means for characterization and imaging. In addition to intensity, phase, coherence, and spectrum variations of scalar light [1], how the vector wave evolves when interacting with a medium for polarized light can reveal the microscopic structure and anisotropy of the medium [2]. The use of polarization optics has recently been expanding rapidly in biomedicine [2-6] and applied in, for example, the characterization of complex random media [7,8], tissue diagnosis [9-12], advanced fluorescence microscopy [13], and clinical and preclinical applications [4,14]. The polarized light interaction with media is succinctly described by the four-by-four Mueller matrix $M$, which transforms the state of polarization (the Stokes vector) of the incident beam to that of the outgoing beam [15]. The medium microscopic structure is encoded in the sixteen elements of the Mueller matrix without a direct link. It requires, therefore, decomposition to interpret the polarized light-medium interaction and link the evolution of the vector light wave to the physical property of a complex medium. The standard polar decomposition (or the Lu-Chipman decomposition [16]) factors the Mueller matrix $M$ into a product of a depolarizer matrix $M_\Delta$, a retarder matrix $M_R$, and a diattenuator matrix $M_D$ where the individual matrices describe anisotropic depolarization, phase, and amplitude modulation of polarized light by the medium, providing a straightforward phenomenolgyical interpretation of basic optical properties (depolarization, birefringence, and dichroism) as well as the underlying microstructure of the medium. Variants of the polar decomposition [17,18] and differential Mueller matrix decomposition [19,20] have also emerged recently. However, no existing decomposition method has accounted for the unique requirement for the decomposition of the Mueller matrices measured inside a backward geometry, which is ubiquitous and often preferred in diverse applications from remote sensing to tissue characterization. In reflection geometry, the probing polarized light traverses the sample along the forward and backward paths in sequence. More importantly, the anisotropic polarization property of the sample seen by the probing beam in the forward and backward paths is reciprocal to each other and not the same in general.

In this article, we present reciprocal polarization imaging of complex media by introducing reciprocal polar decomposition of backscattering Mueller matrices (see Fig.1). In contrast to the Lu-Chipman decomposition of the Mueller matrix into the product of $M_\Delta M_R M_D$, the reciprocal polar decomposition factors the backscattering Mueller matrix into a product of $M_D^{\#} M_R^{\#} M_\Delta M_R M_D$ with the diattenuation and retardance along the backward path specified by the reciprocal of their counterparts $M_D^{\#}$ and $M_R^{\#}$ along the forward path based on the reciprocity of the optical wave in its forward and backward paths. We then demonstrate reciprocal polarization imaging in various applications for quantifying complex non-chiral and chiral media (birefringence resolution target, tissue sections, and glucose suspension) with remarkable clarity and accuracy. We also present types of complex media that Lu-Chipman and differential decompositions of backscattering Mueller matrices lead to erroneous medium polarization properties, whereas reciprocal polar decomposition recovers properly. Reciprocal polarization imaging will be instrumental in imaging complex media and opening new applications of polarization optics in reflection geometry.

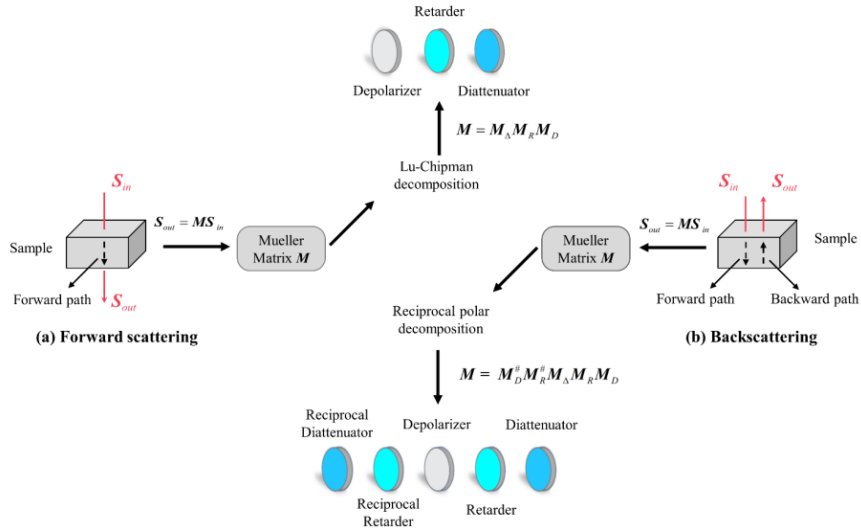

**Fig. 1.** (a) Lu-Chipman decomposition of a forward scattering Mueller matrix. (b) Reciprocal polar decomposition of a backscattering Mueller matrix in reciprocal polarization imaging.

## 2 Reciprocal polarization imaging of complex media

Reciprocal polarization imaging relies on the reciprocity of optical systems in the absence of a magnetic field. We will denote the reciprocal matrix for a Mueller matrix $M$ as $M^{\#}$ for the optical system with the incident and outgoing beams interchanged. The reciprocal Mueller matrix can be written as $M^{\#} = QM^T Q$ where $Q \equiv \text{diag}(1,1,-1,1)$ [21]. We introduce the reciprocal polar decomposition that decomposes a backscattering Mueller matrix into

$$M = M_D^{\#} M_R^{\#} M_\Delta M_R M_D \tag{1}$$

where $M_D$, $M_D^{\#}$ and $M_R$, $M_R^{\#}$ are, respectively, pairs of diattenuator matrices and retarder matrices in the forward and backward paths, and $M_\Delta$ is the depolarizer matrix. The diattenuator and retarder matrices, $M_D$ and $M_R$, take the same form as in Lu-Chipman decomposition (see Supplementary information). Inside an exact backward geometry, the measured Mueller matrix $M$ is the reciprocal of itself, i.e., $M = QM^T Q$. The depolarizer matrix further belongs to the type of the diagonal form [22,23] and reduces to $M_\Delta = M_{\Delta d} \equiv \text{diag}(d_0, d_1, d_2, d_3)$. The reciprocal polar decomposition of the backscattering Mueller matrix $M$ can be rewritten as:

$$QM = M_D^T M_R^T M_{\Delta d}^{'} M_R M_D \tag{2}$$

in terms of the symmetric matrix $QM$ where $M_{\Delta d}^{'} \equiv QM_{\Delta d} = \text{diag}(d_0, d_1, -d_2, d_3)$. The diattenuator, retarder, and depolarizer matrices $M_D$, $M_R$, and $M_\Delta$ of the sample are then determined by decomposition of Eq. (2) (see Methods).

Backscattering Mueller matrices are typically measured slightly off the normal direction to avoid specular reflection. In reciprocal polar decomposition of experimentally measured backscattering Mueller matrices, we replace $QM$ by $[QM + (QM)^T]/2$ to guarantee its symmetry. For comparison, results from Lu-Chipman and differential decompositions are also shown. However, these decomposition methods cannot be applied directly to the backscattering Mueller matrix as the incident and outgoing beams are defined according to different coordinate systems (see Fig. 1). We replace the backscattering Mueller matrix $M$ by $M_{\text{mirror}} M$ flipping the coordinate system for the outgoing beam to be the same as that of the incident beam before Lu-Chipman and differential decompositions where $M_{\text{mirror}} = \text{diag}(1,1,-1,-1)$.

## 3 Results

*3.1 Reciprocal polarization imaging of a birefringence resolution target*

We first imaged NBS 1963A Birefringence Resolution Target (R2L2S1B, Thorlabs) in both backward and forward geometries with our custom polarization imaging system (see Fig. 2). The target contains a liquid crystal polymer pattern sandwiched between two N-BK7 glass substrates and has minimal diattenuation. The extracted linear retardance and orientation angle as well as depolarization for the target by the reciprocal polar decomposition in the backward geometry and by the Lu-Chipman and differential decompositions in both the backward and forward geometries are compared (see Fig. 3). In addition, the horizontal profiles of the target orientation angle, linear retardance, and depolarization along the white line in Fig. 3 are shown in Fig. 4. The mean and standard deviation of the orientation angle, linear retardance, and depolarization for the birefringent (outlined by a red rectangle) and clear (outlined by a white rectangle) regions of the target measured in the forward and backward geometries are summarized in Table 1. The linear retardance from the Lu-Chipman and differential decompositions of the backscattering Mueller matrix account for both forward and backward paths and are multiplied by 1/2 in Fig. 3 and 4, and Table 1 for better comparison.

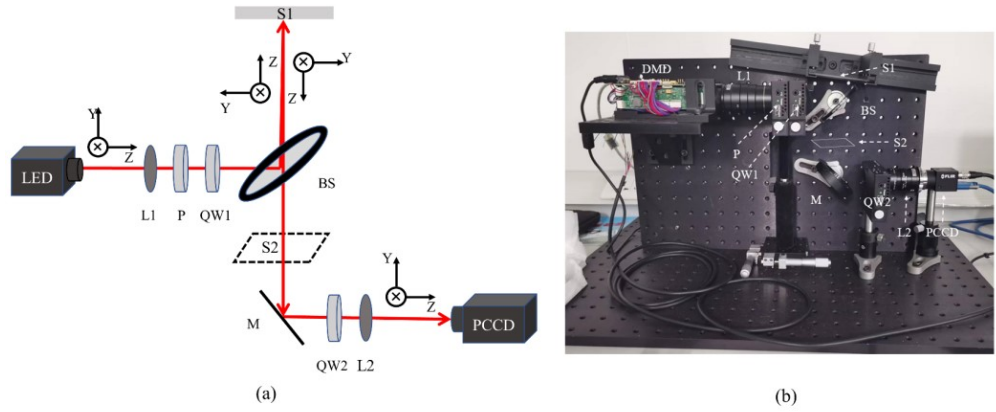

**Fig. 2. The polarization imaging system.** (a) Schematic of the beam path and (b) Photograph of the instrument. P: polarizer; QW: quarter wave plates; L: lens; BS: beam splitter; M: mirror; PCCD: polarization camera; S1: sample position for backscattering Mueller matrix measurement; S2: sample position for forward-scattering Mueller matrix measurement. DMD is set to uniform illumination in experiments.

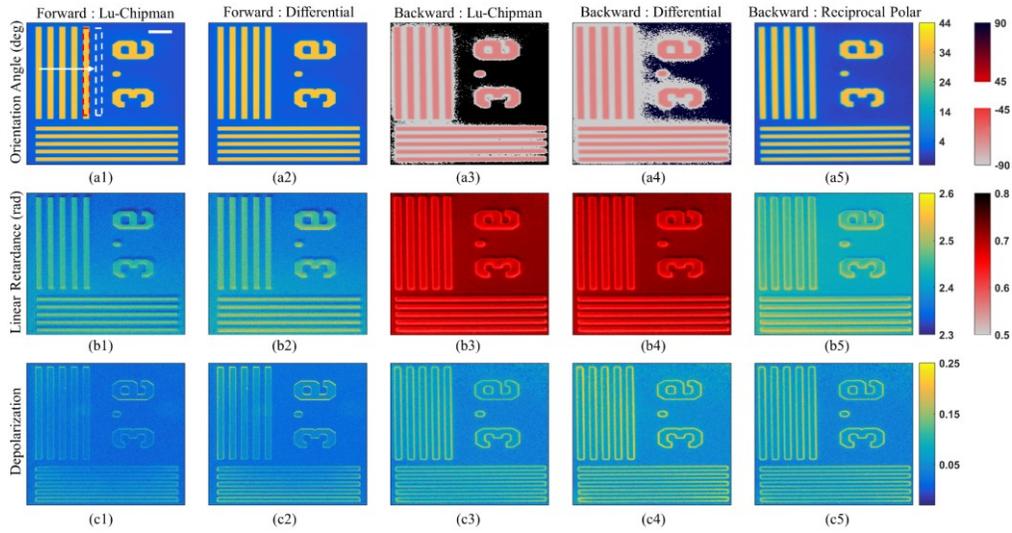

**Fig. 3. Polarization imaging of a birefringence resolution target in forward and backward geometries.** The orientation angle, linear retardance, and depolarization from (a1, b1, c1) Lu-Chipman decomposition and (a2, b2, c2) differential decomposition of the Mueller matrix measured in the forward geometry; (a3, b3, c3) Lu-Chipman decomposition, (a4, b4, c4) differential decomposition, and (a5, b5, c5) reciprocal polar decomposition of the Mueller matrix measured in the backward geometry. The linear retardance (b3, b4) obtained by Lu-Chipman and differential decompositions in the backward geometry has been multiplied by 1/2. Space bar: 0.5 mm.

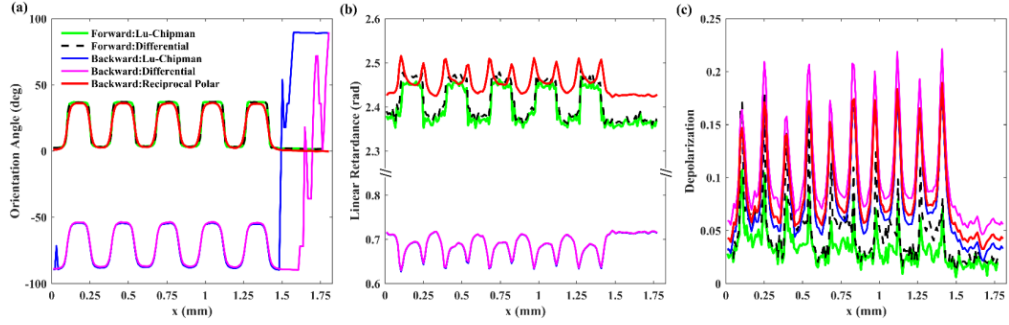

**Fig. 4**. Profiles of (a) the target orientation angle, (b) linear retardance, and (c) depolarization from Lu-Chipman and differential decompositions of the Mueller matrix measured in the forward geometry; Lu-Chipman, differential and reciprocal polar decompositions of the Mueller matrix measured in the backward geometry. The linear retardance (b3, b4) obtained by Lu-Chipman and differential decompositions in the backward geometry has been multiplied by 1/2.

**Table 1. Mean and standard deviation of the orientation angle, linear retardance, and depolarization for the birefringent (subscript "B") and clear (subscript "C") regions of the target measured in the forward and backward geometries. The linear retardance obtained by Lu-Chipman and differential decompositions in the backward geometry has been multiplied by 1/2. Erroneous values are in bold.**

| Parameters | Forward Geometry | | Backward Geometry | | |
|---|---|---|---|---|---|
| | Lu-Chipman | Differential | Lu-Chipman | Differential | Reciprocal |
| $\theta_B$ (deg) | 36.7 (0.6) | 36.7 (0.6) | **-59.9 (7.7)** | **-56.0 (3.6)** | 35.4 (0.5) |
| $\theta_C$ (deg) | 1.7 (0.3) | 1.7 (0.3) | **85.7 (25.5)** | **81.8 (37.0)** | 0.1 (0.3) |
| $\delta_B$ (rad) | 2.45 (0.02) | 2.46 (0.02) | **0.67 (0.01)** | **0.68 (0.01)** | 2.47 (0.02) |
| $\delta_C$ (rad) | 2.37 (0.02) | 2.38 (0.02) | **0.72 (0.01)** | **0.72 (0.01)** | 2.43 (0.01) |
| $\Delta_B$ | 0.03 (0.02) | 0.05 (0.02) | 0.09 (0.04) | 0.10 (0.04) | 0.10 (0.04) |
| $\Delta_C$ | 0.02 (0.02) | 0.02 (0.01) | 0.03 (0.01) | 0.05 (0.01) | 0.04 (0.01) |

The recovered polarization parameters from the reciprocal polar decomposition of the backscattering Mueller matrix are in excellent agreement with those obtained from the Lu-Chipman and differential decompositions of the Mueller matrix measured in the forward geometry other than a stronger depolarization in the former owing to the different detection geometry. The Lu-Chipman and differential decompositions on the backscattering Mueller matrix fail to obtain the correct orientation angle and linear retardance. In particular, the orientation angle obtained by the Lu-Chipman, and differential decompositions of the backscattering Mueller matrix is off by 90 degrees and contains sporadic artifacts. The edges of the birefringent regions exhibit higher retardance and depolarization in the backscattering measurement (see Figs. 3 (b5, c5) and 4). This difference originates from edge diffraction, which leads to increased depolarization and larger retardance of light rays that transverse a longer path within the target. In contrast, the edges of the birefringent regions in the forward geometry only show slightly increased depolarization as the edge diffraction rays are much more dominant in the backscattering geometry where the specular reflection is removed than in the forward geometry. The obtained orientation and linear retardance by the Lu-Chipman and differential decompositions of the forward-scattering Mueller matrix and the reciprocal polar decomposition of the backscattering Mueller matrix agree with the data provided by the manufacturer (see Supplementary information).

### 3.2 Reciprocal polarization imaging of tissue

We then imaged fresh beef sections of thickness 100 μm and 300 μm. The extracted tissue birefringence orientation angle, linear retardance, depolarization, and linear depolarization anisotropy (defined as $A \equiv (|d_1|-|d_2|)/(|d_1|+|d_2|)$ by Lu-Chipman, differential, and reciprocal polar decomposition in the backward geometry are compared to those of the 100-μm

tissue section measured in the forward geometry (see Fig. 5). The mean and standard deviation of the orientation angle, linear retardance, depolarization, and depolarization anisotropy for the whole section and representative fibers are summarized in Tables 2 and 3, respectively. The linear retardance from the Lu-Chipman and differential decompositions of the backscattering Mueller matrix account for both forward and backward paths and are multiplied by 1/2 in Fig. 5 and Tables 2 and 3 for better comparison. The polarization parameter images recovered from the backscattering measurement by different decomposition methods resemble those from the forward geometry in appearance yet with essential differences, allowing inevitable deformation of the tissue section between different measurement geometries.

For the 100 μm tissue section, the orientation angle from the reciprocal polar decomposition is in closer agreement than that computed by the Lu-Chipman and differential decompositions of the backscattering Mueller matrix with the orientation angle measured in the forward geometry (mean squared error: 117 vs. 216 vs. 132, and correlation coefficient: 0.265 vs. 0.206 vs. 0.258). As expected, light depolarization is larger in the backward than in the forward geometry. Moreover, tissue depolarization and depolarization anisotropy are distorted in the Lu-Chipman and differential decompositions of the backscattering Mueller matrices. The images of tissue linear retardance, depolarization, and depolarization anisotropy from the reciprocal polar decomposition are much sharper than those obtained by the Lu-Chipman and differential decompositions of the common backscattering Mueller matrix (sharpness: $5.01 \times 10^3$ vs. $4.71 \times 10^3$ vs. $3.13 \times 10^3$ for tissue linear retardance, $3.64 \times 10^3$ vs. $1.89 \times 10^3$ vs. $7.98 \times 10^2$ for tissue depolarization, and $2.71 \times 10^4$ vs. $2.49 \times 10^3$ vs. $7.25 \times 10^2$ for depolarization anisotropy).

More importantly, Lu-Chipman and differential decomposition in the backward geometry of the 300-μm tissue section produce incorrect orientation angles, retardance, and depolarization anisotropy, whereas the reciprocal polar decomposition succeeds (see Table 2). Furthermore, the images of tissue depolarization from the reciprocal polar decomposition are much sharper than those obtained by the Lu-Chipman and differential decompositions of the common backscattering Mueller matrix (sharpness: $7.32 \times 10^3$ vs. $6.06 \times 10^2$ vs. $3.40 \times 10^2$ for tissue depolarization). Boxplots in Fig. 5 show that the polarization parameters recovered by the reciprocal polar decomposition are the most well-behaved in their data distributions.

For representative fibers highlighted by white arrows in the tissue sections (see Table 3), the orientation angle (38.2 ± 1.1°) from reciprocal polar decomposition of the backscattering Mueller matrix is closer to the expected direction (37.3 ± 0.9°) normal to the fiber than that (44.4 ± 0.8°) obtained by the Lu-Chipman decomposition and (40.1 ± 1.0°) obtained by differential decomposition in the 100-μm section. For the 300-μm section, the orientation angle obtained by the Lu-Chipman and differential decompositions is wrong, yet that by the reciprocal polar decomposition is right. Furthermore, reciprocal polar decomposition shows that the linear retardance is ~ three times for the 300-μm section than that of the 100-μm one, unlike Lu-Chipman and differential decompositions yielding smaller retardance for thicker samples. The depolarization anisotropy of the 100-μm section recovered by the reciprocal polar decomposition in the backward geometry agrees well with that measured in the forward geometry, and depolarization and depolarization anisotropy are larger for the thicker section, as expected. In contrast, the Lu-Chipman and differential decomposition of the backscattering Mueller matrices yields erroneous depolarization anisotropy.

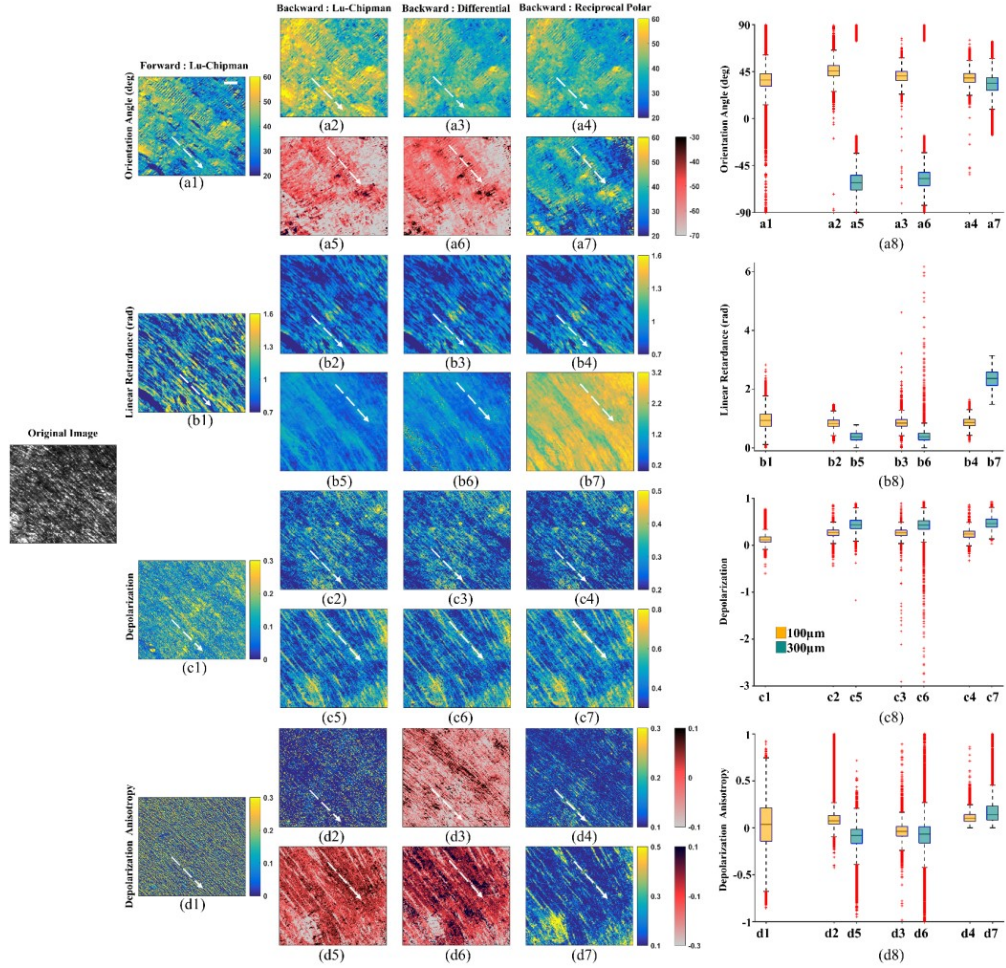

**Fig. 5.** Polarization imaging of fresh beef tissue sections of 100 μm and 300 μm thick. The orientation angle, linear retardance, depolarization, and depolarization anisotropy: (a1, b1, c1, d1) Lu-Chipman decomposition of the Mueller matrix for the 100-μm section measured in the forward geometry; Lu-Chipman decomposition of (a2, b2, c2, d2) the 100-μm section and (a5, b5, c5, d5) the 300-μm section, differential decomposition of (a3, b3, c3, d3) the 100-μm section and (a6, b6, c6, d6) the 300-μm section, and reciprocal polar decomposition of (a4, b4, c4, d4) the 100-μm section and (a7, b7, c7, d7) the 300-μm section measured in the backward geometry. The linear retardance (b2, b3, b5, b6) obtained by Lu-Chipman and differential decompositions in the backward geometry has been multiplied by 1/2. The left inset shows the photo of the sample under unpolarized light illumination. Boxplots are shown for the orientation angle (a8), linear retardance (b8), depolarization (c8), and depolarization anisotropy (d8) of the whole tissue sections obtained by Lu-Chipman, differential, and reciprocal polar decompositions of the Mueller matrices measured in the forward and backward geometries. White arrows highlight representative fibers. Space bar: 0.5 mm.

**Table 2.** Mean and standard deviation of the orientation angle, linear retardance, depolarization, and depolarization anisotropy for the tissue sections (thickness 100 μm vs. 300 μm) measured in the backward geometries compared to those of the 100-μm tissue section measured in the forward geometry. The linear retardance obtained by Lu-Chipman and differential decompositions in the backward geometry has been multiplied by 1/2. Erroneous values are in bold. Gray shading marks contradictory values for 100-um and 300-um tissue sections.

| | Forward Geometry | Backward Geometry | | |
|---|---|---|---|---|
| | Lu-Chipman | Lu-Chipman | Differential | Reciprocal |
| $\theta$ (deg) | 36.7 (11.4) | **45.8 (7.9)** vs **-62.9 (12.6)** | 41.0 (6.8) vs **-58.7 (11.3)** | 39.2 (6.4) vs 33.0 (10.8) |
| $\delta$ (rad) | 0.93 (0.33) | 0.83 (0.17) vs **0.77 (0.30)** | 0.87 (0.18) vs **0.79 (0.52)** | 0.86 (0.17) vs 2.35 (0.32) |
| $\Delta$ | 0.13 (0.10) | 0.26 (0.09) vs 0.44 (0.13) | 0.26 (0.11) vs 0.43 (0.11) | 0.23 (0.10) vs 0.47 (0.12) |
| $A$ | 0.03 (0.27) | 0.11 (0.14) vs 0.10 (0.14) | **-0.13 (0.22)** vs **-0.08 (0.18)** | 0.11 (0.06) vs 0.18 (0.14) |

Table 3. Mean and standard deviation of the orientation angle, linear retardance, depolarization, and depolarization anisotropy for the representative fiber in the tissue sections (thickness 100 µm vs. 300 µm) measured in the backward geometry compared to those of the corresponding fiber in the 100-µm tissue section measured in the forward geometry. The linear retardance obtained by Lu-Chipman and differential decompositions in the backward geometry has been multiplied by 1/2. Erroneous values are in bold. Gray shading marks contradictory values for 100-um and 300-um tissue sections.

|  | Forward Geometry | Backward Geometry | | |
|---|---|---|---|---|
|  | Lu-Chipman | Lu-Chipman | Differential | Reciprocal |
| $\theta$ (deg) | 36.6 (2.1) | **44.4 (0.8)** vs **-55.82 (12.6)** | 40.1 (1.0) vs **–50.9 (5.9)** | 38.2 (1.1) vs 40.9 (5.8) |
| $\delta$ (rad) | 0.80 (0.06) | 0.84 (0.06) vs **0.58 (0.17)** | 0.85 (0.06) vs **0.62 (0.18)** | 0.85 (0.06) vs 2.51 (0.17) |
| $\Delta$ | 0.16 (0.04) | 0.36 (0.06) vs 0.68 (0.05) | 0.36 (0.06) vs 0.68 (0.06) | 0.32 (0.07) vs 0.68 (0.05) |
| $A$ | 0.21 (0.17) | **0.03 (0.03)** vs **-0.19 (0.20)** | -0.21 (0.10) vs -0.28 (0.19) | 0.19 (0.10) vs 0.41(0.17) |

*3.3 Reciprocal polarization imaging of a chiral medium*

Experimentally measured Mueller matrices for a chiral turbid medium (scattering coefficient $\mu_s$ =0.6 mm$^{-1}$, anisotropy factor g=0.91, glucose concentration 5 M, and thickness 10 mm) in the forward and backward scattering geometries reported by Manhas et al. [8] was analyzed to demonstrate the applicability of reciprocal polarization imaging for chiral media. This medium is nonbirefringent, and the apparent linear retardance is attributed to light scattering [8]. The reciprocal polar decomposition shows the increase of linear retardance $\delta$ and optical rotation $\Psi$ at a similar rate in the backward vs. forward scattering geometries, consistent with larger linear retardance and optical rotation in the backward scattering geometry owing to the increased path lengths for the detected photons. In contrast, the Lu-Chipman and differential decomposition of the backscattering Mueller matrix yields an erroneous optical rotation in the backscattering geometry (see Table 4). Furthermore, the much longer path lengths of the detected photons further lead to increased light depolarization recovered from Lu-Chipman, differential, and reciprocal polar decompositions in the backward than forward geometries.

Table 4. Reciprocal polar and differential decomposition of the Mueller matrix measured in the backward geometry for a chiral turbid medium and the extracted polarization parameters compared to the Lu-Chipman decomposition of the Mueller matrices measured in the forward and backward geometries for the same medium. Erroneous values are in bold.

Backscattering $M$:
$$\begin{pmatrix} 1 & -0.115 & -0.066 & 0.023 \\ -0.111 & 0.759 & -0.061 & -0.001 \\ -0.018 & 0.151 & -0.435 & -0.139 \\ -0.046 & 0.006 & 0.128 & -0.334 \end{pmatrix}$$

Symmetrized $(QM + (QM)^T)/2$:
$$\begin{pmatrix} 1 & -0.113 & -0.024 & -0.012 \\ -0.113 & 0.759 & -0.106 & 0.002 \\ -0.024 & -0.106 & 0.435 & 0.134 \\ -0.012 & 0.002 & 0.134 & -0.334 \end{pmatrix}$$

$M_\Delta$:
$$\begin{pmatrix} 0.997 & 0 & 0 & 0 \\ 0 & 0.790 & 0 & 0 \\ 0 & 0 & -0.425 & 0 \\ 0 & 0 & 0 & -0.359 \end{pmatrix}$$

$M_R$:
$$\begin{pmatrix} 1 & 0 & 0 & 0 \\ 0 & 0.952 & -0.305 & -0.035 \\ 0 & 0.307 & 0.937 & 0.166 \\ 0 & -0.018 & -0.168 & 0.986 \end{pmatrix}$$

$M_D$:
$$\begin{pmatrix} 1 & -0.066 & -0.020 & -0.013 \\ -0.066 & 1 & 0.001 & 0.000 \\ -0.020 & 0.001 & 0.998 & 0.000 \\ -0.013 & 0.000 & 0.000 & 0.998 \end{pmatrix}$$

| Parameters | Forward Geometry | Backward Geometry | | |
|---|---|---|---|---|
|  | Lu-Chipman | Lu-Chipman | Differential | Reciprocal |
| $\Delta$ | 0.033 | 0.473 | 0.478 | 0.475 |
| $\delta$ (rad) | 0.055 | 0.336 | 0.339 | 0.170 |
| $\Psi$ (rad) | 0.069 | **0.034** | **0.036** | 0.157 |

*3.4 Comparison of the performance of Lu-Chipman, differential, and reciprocal polar decompositions of backscattering Mueller matrices*

The fundamental difference between polarization imaging in the forward or backscattering geometry is that the anisotropic property of the medium observed by the probing beam along the forward path and backward path is not identical but reciprocal. Unfortunately, the well-known Lu-Chipman and differential decomposition methods do not account for this distinctive property of backscattering polarimetry.

Only reciprocal polar decomposition can recover the optical rotations properly, whereas Lu-Chipman and differential decompositions produce erroneous results from backscattering Mueller matrices (see Sec. 3.3). For specimens of negligible optical rotation, the polarization parameters obtained from all three decompositions may look similar but have significant discrepancies. For non-chiral media with linear retardance less than $\pi/2$, Lu-Chipman and differential decompositions yield similar orientation angles, double retardance (compared with forward Lu-Chipman decomposition and backward reciprocal polar decomposition), and incorrect depolarization anisotropy (see Fig. 5, Table 2 and 3) compared to reciprocal polar decomposition. The linear retardance from Lu-Chipman and differential decompositions double because they count forward and backward paths together [24]. For non-chiral media with retardance exceeding $\pi/2$, Lu-Chipman and differential decompositions in backward configuration produce incorrect orientation angles, retardance, and depolarization anisotropy (see Fig. 3, 4, and 5, Tables 1, 2, and 3). In addition, although the depolarization obtained by all three decompositions is similar in backward geometry other than observable sharpness difference, the depolarization anisotropy, which marks the local anisotropic structure, is dissimilar. Depolarization anisotropy reveals significant structural features, particularly for media with high depolarization. Reciprocal polar decomposition yields the sharpest depolarization and depolarization anisotropy images that resemble those from the Lu-Chipman decomposition measured in forward geometry. Finally, reciprocal polar decomposition is the only method that correctly reveals that the linear retardance increases with the sample thickness from 100 μm to 300 μm.

## 4 Discussion and Conclusion

The Mueller matrix decomposition plays a significant role in interpreting the polarized light-medium interaction and relating the evolution of the vector light wave to the physical property of a complex medium. Similar to polarization imaging that relies on the Lu-Chipman decomposition for factoring the forward-scattering Mueller matrix into a product of three matrices describing the medium depolarization, birefringence, and dichroism, reciprocal polarization imaging achieves the feat by the reciprocal polar decomposition of Mueller matrices measured in the backward scattering geometry. It factors the backscattering Mueller matrix into a product of $M_D^\# M_R^\# M_\Delta M_R M_D$ with the diattenuation and retardance along the backward path specified by the reciprocal of their counterparts $M_D^\#$ and $M_R^\#$ along the forward path based on the reciprocity of the optical wave in the forward and backward paths. Although reciprocal decomposition may look similar to symmetric decomposition, they differ fundamentally and should not be confused. Reciprocal polar decomposition strictly enforces reciprocity and can only be applied to backscattering polarimetry. It contains ten degrees of freedom as the backscattering Mueller matrices are inherently constrained ($QM$ is a symmetric matrix). Symmetric decomposition, on the other hand, is a generic Mueller matrix decomposition method and has sixteen degrees of freedom. It does not account for the reciprocity of the backward and forward paths in backscattering measurements. Symmetric decomposition was found to suffer from poor characterization of complex media in reflection geometry (see Supplementary information). Despite being a phenomenological theory, the reciprocity assumed in reciprocal polar decomposition strictly holds when single-scattering dominates and is justified in multiple scattering situations as long as the angular distribution of the incident photons and the bounced-back photons is close to each other surrounding the optical axis at the plane where photons are backscattered.

Compared to the forward-scattering Mueller matrices, the backscattering Mueller matrices have reduced degrees of freedom of ten. These ten degrees of freedom exactly map to ten polarization parameters: the diattenuation vector **D**, the retardance vector (i.e., the linear retardance $\delta$, its orientation $\theta$, and the optical rotation $\Psi$), and the depolarization factors $d_0$, $d_1$, $d_2$, and $d_3$. Reciprocal polarization imaging realizes a straightforward interpretation of the polarization measurement in the backward geometry and provides the determination of the same set of ten polarization parameters typically extracted from Mueller matrices measured in the forward geometry. It presents a significant advantage as the backward geometry is often preferred or the only feasible approach for bulk samples and is more convenient in applications.

We have demonstrated reciprocal polarization imaging with various complex non-chiral and chiral media and have shown the superiority of the reciprocal polar decomposition to the Lu-Chipman and differential decompositions of backscattering Mueller matrices. The Lu-Chipman and differential decompositions of the backscattering Mueller matrices produce similar yet distorted retardance, depolarization, and depolarization anisotropy images compared to the reciprocal polar decomposition for media of a low retardance. In particular, the Lu-Chipman and differential decompositions fail to obtain the correct depolarization anisotropy from the backscattering Mueller matrices, whereas the reciprocal polar decomposition succeeds. For media of retardance exceeding $\pi/2$, significant errors in the orientation angle and retardance are further observed in the Lu-Chipman and differential decompositions of the backscattering Mueller matrices. Additional examples of the failure of Lu-Chipman decomposition of backscattering Mueller matrices are given in Supplementary information. In all cases, the polarization properties of complex media determined by the reciprocal polar decomposition of the backscattering Mueller matrices are in excellent agreement with those obtained by polarization imaging of the same sample in the forward geometry. Furthermore, when light depolarization ($\Delta$) is elevated, the accuracy of recovered media polarization parameters deteriorates from polarization imaging [25]. The performance of reciprocal polar decomposition for complex media of high depolarization warrants further study and will be published later.

Polarization imaging with Mueller matrix decomposition is a powerful means to quantify the diattenuation, retardance, and depolarization of complex media and link to the underlying microstructure and anisotropy. The measurement of the spatially resolved polarization properties has a wide array of applications in non-invasive characterization and diagnosis of complex random media [7,8], such as biological cells and tissue [9-12], for clinical and preclinical applications [4,14]. However, such measurement has traditionally been limited to transmission geometry. As backward geometry is ubiquitous and often preferred in applications from remote sensing to biomedical imaging, reciprocal polarization imaging will be instrumental in imaging complex media and opening new applications of polarization optics in reflection geometry.

## 5 Materials and Methods

*5.1 Reciprocal polar decomposition*

The decomposition for Eq. (2) is performed similar to the symmetric decomposition [18] but much more straightforwardly. Noting

$$\left(1 - D^2\right) \boldsymbol{M}_D^{-1} = \boldsymbol{G} \boldsymbol{M}_D \boldsymbol{G} \tag{3}$$

when the Mueller matrix **M** does not represent a perfect analyzer $(D < 1)$ [16] where $\boldsymbol{G} \equiv \operatorname{diag}(1, -1, -1, -1)$ is the Minkowski metric matrix, we have

$$\left(\boldsymbol{Q}\boldsymbol{M}\boldsymbol{G}\right)\left(\boldsymbol{M}_D \boldsymbol{G}\right) = \left(1 - D^2\right) \boldsymbol{M}_D^T \boldsymbol{M}' \tag{4}$$

with

$$M' \equiv M_R^T M'_{\Delta d} M_R = \begin{pmatrix} d_0 & \mathbf{0}^T \\ \mathbf{0} & m' \end{pmatrix} \qquad (5)$$

Expanding Eq. (4) after substituting Eqs. (S2) and (5) for $M_D$ and $M'$, respectively, lead to

$$QMG \begin{pmatrix} 1 \\ \mathbf{D} \end{pmatrix} = d_0 (1 - D^2) \begin{pmatrix} 1 \\ \mathbf{D} \end{pmatrix} \qquad (6)$$

This equation has one unique positive eigenvalue ($= d_0 (1 - D^2)$) and the associated eigenvector[18] providing the diattenuation vector $\mathbf{D}$ satisfying $|\mathbf{D}| = D < 1$. With the determination of the diattenuation vector $\mathbf{D}$, the diattenuator matrix $M_D$ is obtained.

Afterward, Eq. (2) reduces to

$$M_R^T M'_{\Delta d} M_R = N \equiv M_D^{-1} Q M M_D^{-1} \qquad (7)$$

and equivalently

$$m_R^T \mathrm{diag}(d_1, -d_2, d_3) m_R = n \qquad (8)$$

for the bottom right 3×3 submatrices $m_R$ and $n$ of the matrices $M_R$ and $N$. An orthogonal decomposition on $n$ can then be used to determine $m_R$ and $d_i (i = 1, 2, 3)$. In addition, the order and the sign of the eigenvectors should be determined with *priori* information, if available, regarding the depolarization properties of the medium. A convention of ordering the eigenvectors, which may be multiplied by $\pm 1$, to have a minimal total retardance is adopted in the absence of *a priori* information, similar to the Lu-Chipman and symmetric decompositions[16,26]. The reciprocal polar decomposition of the backscattering Mueller matrix $M$ into products (1) and (2) is then obtained.

We note that the degeneracy of the eigenvalues, $d_1$, $-d_2$, and $d_3$, will render either part or complete loss of the determination of the retardance parameters as an arbitrary rotation can be applied to the degenerate eigenvectors. One potential degeneracy in the reciprocal polar decomposition arises for a depolarizer matrix $M_{\Delta d} = \mathrm{diag}(d_0, d_1, -d_1, d_3)$, which is fortunately rare and will result in an indeterminate circular retardance.

*5.2 Polarization imaging system*

The polarization imaging system is shown in Fig. 2. Collimated light ($\lambda = 633$ nm) passes through a polarization state generator consisting of a rotating polarizer and a rotating quarter-wave plate and illuminates the sample after reflection by a beam splitter. The backscattered light by the sample is reflected by a mirror and passes through a second rotating quarter-wave plate before being recorded by a polarization camera (BFS-U3-51S5P-C, FLIR). The second rotating quarter-wave plate and the linear polarizers along 0°, 45°, 90°, and 135° directions of the polarization camera form the polarization state analyzer. The Mueller matrix for the sample is measured in both the backward (sample at S1) and forward (sample at S2 and an additional mirror at S1) geometries after the removal of the stray light contributions and calibration of the Mueller matrix of the imaging system itself (see Supplementary information). In the backward geometry, the specular reflection from the sample surface is removed by slightly tilting the sample surface (~ 5°). This Mueller imaging system has a field of view of 1.8×2.0 cm². The acquisition of one complete Mueller matrix of the sample takes one minute.

## 5.3 Materials

Silverside beef rich in muscle fibers was selected for the experiment. One $1\times1\times1$ cm$^3$ tissue was mounted in OCT embedding compound (Sakura) and kept at $-80$ °C for 10 hours. Tissue sections were cut using a cryostat microtome (HM525, Thermofisher) and mounted on a cover slip.

**Funding.** Natural Science Foundation of Zhejiang Province (LZ16H180002); National Natural Science Foundation of China (61905181); Wenzhou Municipal Science and Technology Bureau (ZS2017022); National Science Foundation of U.S. (1607664).

**Acknowledgments.** We thank for Lili Ma's assistance in preparing fresh beef section samples.

**Disclosures.** The authors declare no conflicts of interest.

**Data availability.** Data underlying the results presented in this paper are not publicly available at this time but may be obtained from the authors upon reasonable request.

**Supplemental document.** See Supplement 1 for supporting content.

# Reciprocal polarization imaging of complex media: supplemental document

Zhineng Xie, Guowu Huang, Weihao Lin, Xin Jin, Xiafei Qian, Yifan Ge, Yansen Hu and Min Xu[*]

[*]Corresponding author. Email: minxu@hunter.cuny.edu

**This file includes:**
    Figs. S1 to S9
    Table S1 to S6

## 1. Lu-Chipman polar decomposition

The Lu-Chipman decomposition of Mueller matrices is briefly outlined here for clarity and comparison with the reciprocal polar decomposition. Lu-Chipman decomposition decomposes the Mueller matrix into a product of three matrices: a depolarizer matrix $M_\Delta$, a retarder matrix $M_R$, and a diattenuator matrix $M_D$ [1,2], i.e.,

$$M = M_\Delta M_R M_D \tag{S1}$$

Different order for the above product has been introduced, and Eq. (S1) is widely adopted because the family in which the diattenuator matrix comes ahead of the depolarizer matrix always leads to a physically realizable Mueller matrix [3]. Here the diattenuator matrix is a symmetric matrix given by

$$M_D = \begin{pmatrix} 1 & D^T \\ D & m_D \end{pmatrix} \tag{S2}$$

in which $m_D = \sqrt{1-D^2}\, I + \left(1-\sqrt{1-D^2}\right)\hat{D}\hat{D}^T$, $I$ is a $3 \times 3$ identity matrix, $\hat{D}$ is the unit vector for the diattenuation vector $D = M_{00}^{-1}(M_{01}, M_{02}, M_{03})^T$ of length $D$, and $M_{ij}$ ($i,j = 0,1,2,3$) is the element at the $i$-th row and $j$-th column of the Mueller matrix $M$. The length is assumed $D < 1$ to ensure a non-singular matrix $M_D$ (for $M_D^{-1}$ to exist). The length $D = 1$ corresponds to a perfect analyzer that $M$ has an indeterminate decomposition[1].

The retarder matrix is written as

$$M_R = \begin{pmatrix} 1 & 0^T \\ 0 & m_R \end{pmatrix} \tag{S3}$$

where $m_R$ is a positive definite three-dimensional rotation matrix. The retarder matrix $M_R$ can further be expressed as an ordered product of a circular retarder of retardance $\Psi$ and a linear retarder of retardance $\delta$ with its fast axis oriented at $\theta$. Their values are given by

$$\delta = \cos^{-1}\left( \sqrt{\left((M_R)_{1,1} + (M_R)_{2,2}\right)^2 + \left((M_R)_{1,2} - (M_R)_{2,1}\right)^2} - 1 \right) \tag{S4}$$

$$\theta = -\frac{1}{2}\mathrm{atan2}\left((M_R)_{1,3}, (M_R)_{2,3}\right) \tag{S5}$$

and

$$\Psi = \frac{1}{2}\mathrm{atan2}\left((M_R)_{1,2} - (M_R)_{2,1}, (M_R)_{1,1} + (M_R)_{2,2}\right) \tag{S6}$$

Furthermore, the total retardance and the depolarization coefficient are given by

$$R = \cos^{-1}\left[2\cos^2\Psi\cos^2\left(\frac{\delta}{2}\right) - 1\right] \tag{S7}$$

and

$$\Delta = 1 - \frac{1}{3}\left[\text{tr}\left|\frac{M_\Delta}{(M_\Delta)_{00}}\right| - 1\right] \tag{S8}$$

## 2. Differential decomposition

Differential decomposition assumes of the medium polarization properties is uniform along the optical path [4] and the derivative of the Mueller matrix along the optical path, z, can be written as

$$\frac{dM(z)}{dz} = m(z)M(z) \tag{S9}$$

where the matrix $m$ contains the elementary polarimetric properties of the sample per unit of distance (i.e. specific properties). The matrix $m$ can be expressed as

$$m = \begin{pmatrix} \alpha & \beta & \gamma & \partial \\ \beta' & \alpha_1 & \mu & -\upsilon \\ \gamma' & -\mu' & \alpha_2 & \eta \\ \partial' & \upsilon' & -\eta' & \alpha_3 \end{pmatrix} \tag{S10}$$

Here $\beta$ and $\gamma$ is the linear dichroism along x-y (horizontal-vertical) and ± 45° laboratory axes, respectively; $\partial$ is the circular dichroism; $\eta$ and $\upsilon$ are linear birefringence along x-y and ± 45°axes, respectively; $\mu$ is the circular birefringence; $\alpha$ and $\alpha_1, \alpha_2, \alpha_3$, respectively are the isotropic absorption and the anisotropic absorptions (or depolarizations) along x-y, ±45° and circular axes, respectively. Note that in absence of depolarization, $\alpha = \alpha_1 = \alpha_2 = \alpha_3 = 0$ [5].

If the medium properties (both polarizing and depolarizing) are uniformly distributed along the propagation direction z, i.e., the differential matrix $m$ is z-independent, Eq. (S9) can be integrated to yield

$$L = \ln M = md = \begin{pmatrix} l_{00} & l_{01} & l_{02} & l_{03} \\ l_{10} & l_{11} & l_{12} & l_{13} \\ l_{20} & l_{21} & l_{22} & l_{23} \\ l_{30} & l_{31} & l_{32} & l_{33} \end{pmatrix} \tag{S11}$$

with $L$ being the matrix logarithm of $M$ and $d$ being the optical path length. $L$ can be further decomposed into two terms as $L = L_m + L_u$ [6,7], where the G-antisymmetric and G-symmetric matrices $L_m$ and $L_u$ are given by:

$$L_m = \frac{1}{2}(L - GL^TG) = \frac{1}{2}\begin{pmatrix} 0 & l_{01}+l_{10} & l_{02}+l_{20} & l_{03}+l_{30} \\ l_{01}+l_{10} & 0 & l_{12}-l_{21} & l_{13}-l_{31} \\ l_{02}+l_{20} & l_{21}-l_{12} & 0 & l_{23}-l_{32} \\ l_{03}+l_{30} & l_{31}-l_{13} & l_{32}-l_{23} & 0 \end{pmatrix} \tag{S12}$$

$$L_u = \frac{1}{2}(L + GL^TG) = \frac{1}{2}\begin{pmatrix} 2l_{00} & l_{01}-l_{10} & l_{02}-l_{20} & l_{03}-l_{30} \\ l_{01}-l_{10} & 2l_{11} & l_{12}+l_{21} & l_{13}+l_{31} \\ l_{02}-l_{20} & l_{21}+l_{12} & 2l_{22} & l_{23}+l_{32} \\ l_{03}-l_{30} & l_{31}+l_{13} & l_{32}+l_{23} & 2l_{33} \end{pmatrix} \tag{S13}$$

in which $G=\text{diag}(1,-1,-1,-1)$ is the Minkowski metric matrix.

The accumulated polarization parameters [5,6] can be determined from the elements of the $L_m$ and $L_u$ matrices as

$$\delta = \sqrt{(L_m)_{1,3}^2 + (L_m)_{2,3}^2} \quad (S14)$$

$$\theta = \frac{1}{2}\operatorname{atan2}(-(L_m)_{1,3}, (L_m)_{2,3}) \quad (S15)$$

$$\Psi = \frac{(L_m)_{1,2}}{2} \quad (S16)$$

$$\Delta = 1 - \frac{d_1 + d_2 + d_3}{3} \quad (S17)$$

$$d_i = e^{(L_u)_{ii}} \ (i = 1,2,3) \quad (S18)$$

## 3. Calibration of the polarization imaging system

The polarization state generator (consisting of a rotating polarizer and a rotating quarter wave plate) generates four input polarization states: $0°$ (Stokes vector $[1\ 1\ 0\ 0]^T$), $45°$ (Stokes vector $[1\ 0\ 1\ 0]^T$), $90°$ (Stokes vector $[1\ -1\ 0\ 0]^T$) linear polarization, and circular polarization (Stokes vector $[1\ 0\ 0\ 1]^T$) states. The polarization state analyzer of a rotating quarter wave plate and a polarization camera[8] uses the approach presented.

The calibration of the Mueller imaging system follows the steps outlined in Fig. S1. First, Stokes vector $S_{0°}$, $S_{45°}$, $S_{90°}$, and $S_{circular}$ for the four input polarization states were measured using the configuration (a). Second, Mueller matrices of mirrors M1 and M2, $M_{M1}$ and $M_{M2}$, were measured using the configuration (b). Third, the reflection and transmission Mueller matrices of the beam splitter BS, $R$ and $T$, were measured in the configuration (c) and (d), respectively. The obtained values are given below:

$$\begin{aligned}
S_{0°} &= [1\ \ 0.983\ \ 0.002\ \ -0.025]^T \\
S_{45°} &= [0.984\ \ 0.019\ \ 0.965\ \ 0.020]^T \\
S_{90°} &= [0.978\ \ -0.952\ \ 0.003\ \ -0.021]^T \\
S_{circular} &= [0.972\ \ 0.03\ \ 0.061\ \ 0.947]^T
\end{aligned} \quad (S18)$$

$$S_{in} = \begin{bmatrix} S_{0°} & S_{90°} & S_{45°} & S_{circular} \end{bmatrix} \quad (S19)$$

$$M_{M1} = \begin{bmatrix} 1 & 0.141 & 0.013 & -0.005 \\ 0.135 & 0.986 & 0.001 & 0.033 \\ -0.020 & 0.059 & -0.997 & -0.028 \\ -0.032 & -0.028 & 0.068 & -0.999 \end{bmatrix} \quad (S20)$$

$$M_{M2} = \begin{bmatrix} 1 & -0.053 & -0.027 & -0.009 \\ -0.052 & 0.9956 & -0.025 & -0.021 \\ 0.038 & -0.002 & -0.942 & 0.371 \\ -0.002 & -0.073 & -0.380 & -0.884 \end{bmatrix} \quad (S21)$$

$$R = \begin{bmatrix} 1 & 0.366 & -0.011 & -0.024 \\ 0.344 & 0.998 & -0.078 & -0.038 \\ -0.031 & -0.013 & -0.851 & -0.316 \\ -0.033 & -0.034 & 0.281 & -0.814 \end{bmatrix} \quad (S22)$$

$$T = \begin{bmatrix} 1 & -0.395 & 0.015 & -0.001 \\ -0.378 & 0.997 & 0.006 & -0.023 \\ 0.025 & -0.050 & 0.907 & 0.103 \\ 0.032 & -0.001 & -0.045 & 0.874 \end{bmatrix} \quad (S23)$$

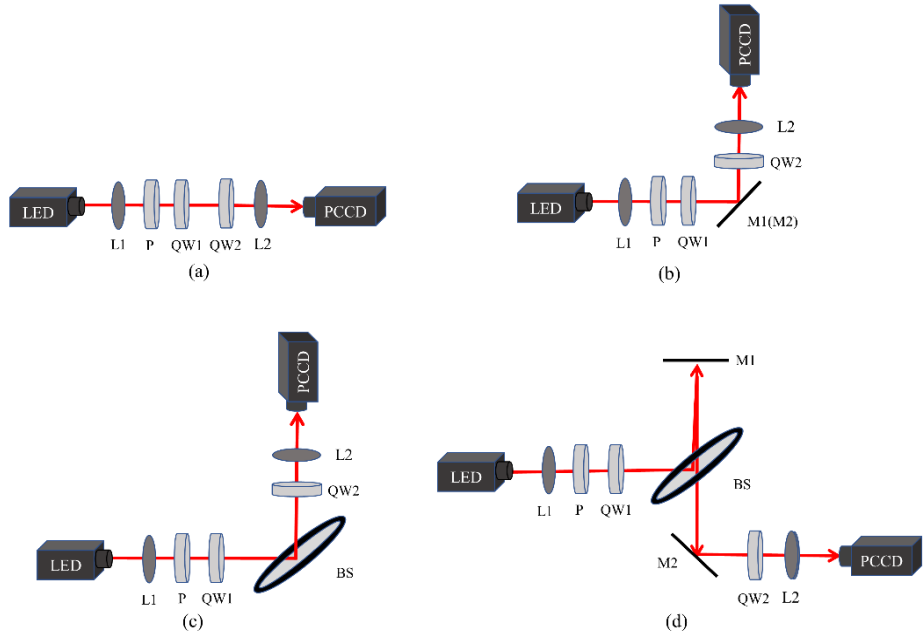

Fig. S1. Calibration of the polarization imaging system.

The measured Mueller matrix $M_{out}$ is given by
$$M_{out} = M_{M2} T M_{Sample} R S_{in} \quad (S24)$$
in the backscattering geometry and is given by
$$M_{out} = M_{M2} M_{Sample} T M_{M1} R S_{in} \quad (S25)$$
in the forward geometry, which contains an additional mirror M1 placed at the S1 position. The Mueller matrix $M_{sample}$ for the sample is then solved from $M_{out}$.

## 4. Birefringent resolution target
The NBS 1963A Birefringence Resolution Target (Thorlabs) has 26 groups of lines, with five horizontal lines and five vertical lines in each group. The line group of 3.6 cycles/mm with a minimum period of 0.278 mm is selected for imaging in the experiment (see Fig. S2).

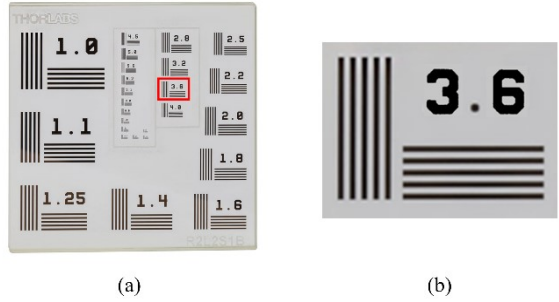

Fig. S2. NBS 1963A Birefringence Resolution Target.

The polarization parameters of the selected horizontal line group provided by the manufacturer are summarized in Table S1 when the target is oriented as in the main text.

**Table S1. The polarization parameters provided by the manufacturer.**

| $\theta_B$ (deg) | $\theta_C$ (deg) | $\delta_B$ (rad) | $\delta_C$ (rad) |
|---|---|---|---|
| 34.5 ± 4.3 | 0.1 ± 3.8 | 2.436 ± 0.084 | 2.383 ± 0.068 |

## 5. Measured orientation angle, linear retardance, and depolarization of the birefringent resolution target when placed in different directions

Measurements were performed for the target when placed in three different directions (#1: the horizontal line along the *x*-axis, #2: the target rotated 14.3° counterclockwise from #1, #3: the target rotated 13.3° clockwise from #1). The orientation angle, linear retardance, and depolarization of the target for #1, #2, and #3 from the Lu-Chipman decomposition of the forward scattering Mueller matrix, the Lu-Chipman decomposition, and the reciprocal polar decomposition of the backscattering Mueller matrix were obtained.

The orientation angle (θ), linear retardance (δ), and depolarization (Δ) are shown in Figs. S3-S5. Their mean and standard deviation are shown in Table S2-S4. The orientation angles and the linear retardance obtained from the Lu-Chipman decomposition of the forward-scattering Mueller matrix and the reciprocal polar decomposition of the backscattering Mueller matrix are in excellent agreement. The orientation angles of the target for #1, #2, and #3 obtained by both the Lu-Chipman decomposition in the forward geometry and the reciprocal polar decomposition in the backward direction correctly follow their relative rotations. As expected, the depolarization of the backscattering Mueller matrix is larger than that of the forward scattering Mueller matrix.

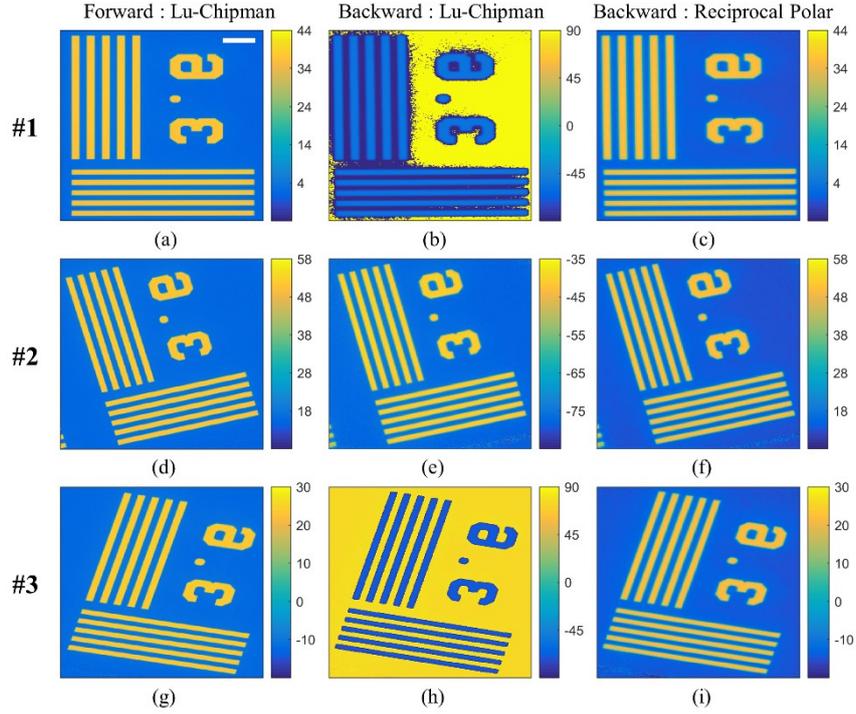

Fig. S3. Orientation angle (in degrees) of the target placed in different directions. Space bar: 0.5 mm.

Table S2. Mean and standard deviation of the orientation angle ($\theta$). $\theta_B$: Orientation angle of the birefringent region. $\theta_C$: Orientation angle of the clear region.

|  |  | Forward Geometry | Backward Geometry | |
|---|---|---|---|---|
|  |  | Lu-Chipman | Lu-Chipman | Reciprocal Polar |
| #1 | $\theta_B$ (deg) | 36.67 (0.59) | -59.85 (7.70) | 35.35 (0.49) |
|  | $\theta_C$ (deg) | 1.67 (0.33) | 85.67 (25.46) | 0.10 (0.28) |
| #2 | $\theta_B$ (deg) | 50.81 (1.53) | -41.05 (0.92) | 49.07 (1.10) |
|  | $\theta_C$ (deg) | 16.27 (0.33) | -76.84 (0.33) | 14.12 (0.29) |
| #3 | $\theta_B$ (deg) | 23.47 (1.64) | -69.96 (0.82) | 20.83 (0.90) |
|  | $\theta_C$ (deg) | -11.56 (0.32) | 76.05 (0.31) | -13.78 (0.27) |

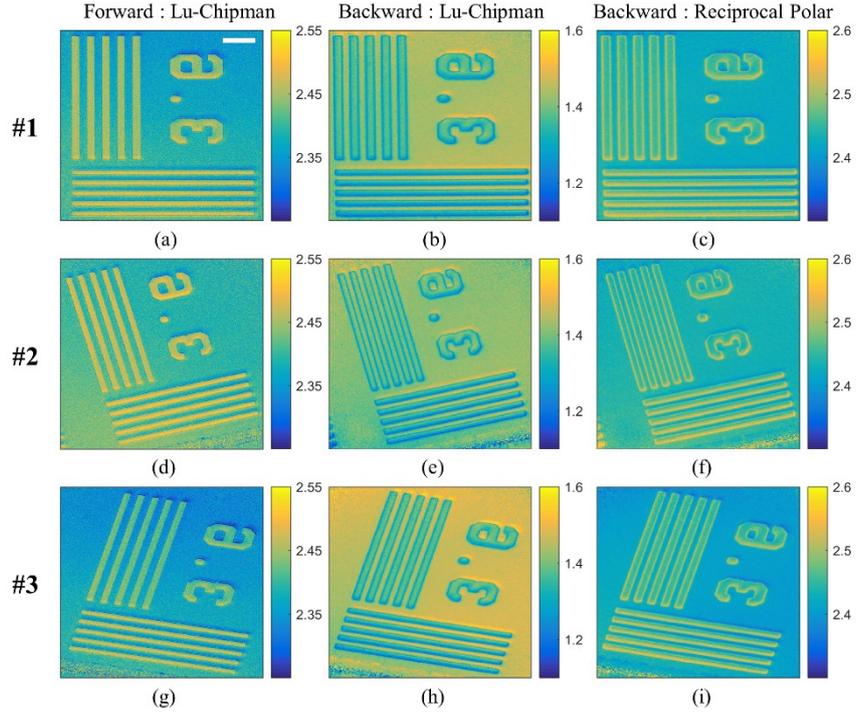

Fig. S4. Linear retardance (rad) of the target placed in different directions. Space bar: 0.5 mm.

Table S3. Mean and standard deviation of linear retardance (rad). $\delta_B$: Linear retardance of the birefringent region. $\delta_C$: Linear retardance of the clear region.

|  |  | Forward Geometry | Backward Geometry | |
|---|---|---|---|---|
|  |  | Lu-Chipman | Lu-Chipman | Reciprocal Polar |
| #1 | $\delta_B$ (rad) | 2.448 (0.017) | 1.344 (0.038) | 2.471 (0.019) |
|  | $\delta_C$ (rad) | 2.370 (0.024) | 1.436 (0.010) | 2.429 (0.005) |
| #2 | $\delta_B$ (rad) | 2.465 (0.019) | 1.353 (0.045) | 2.467 (0.022) |
|  | $\delta_C$ (rad) | 2.396 (0.021) | 1.417 (0.011) | 2.433 (0.005) |
| #3 | $\delta_B$ (rad) | 2.419 (0.022) | 1.354 (0.034) | 2.466 (0.017) |
|  | $\delta_C$ (rad) | 2.348 (0.024) | 1.467 (0.011) | 2.407 (0.005) |

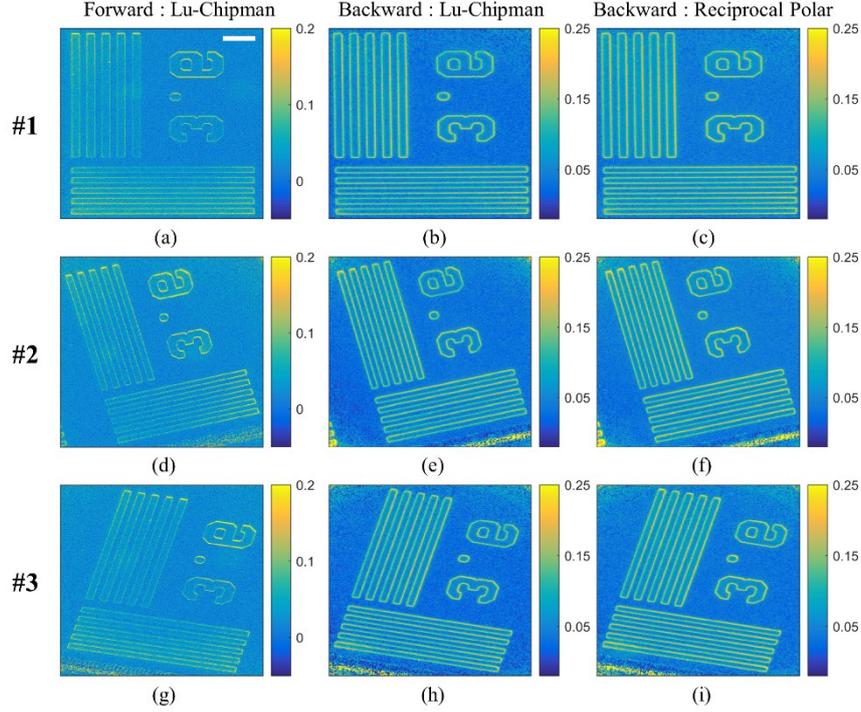

Fig.S5. Depolarization of the target placed in different directions. Space bar: 0.5 mm.

Table S4. Mean and standard deviation of depolarization of the target. $\Delta_B$: Depolarization of the birefringent region. $\Delta_C$: Depolarization of the clear region.

|    |            | Forward Geometry | Backward Geometry | |
|----|------------|------------------|-------------------|---|
|    |            | Lu-Chipman       | Lu-Chipman        | Reciprocal Polar |
| #1 | $\Delta_B$ | 0.032 (0.018)    | 0.093 (0.036)     | 0.100 (0.036) |
|    | $\Delta_C$ | 0.016 (0.015)    | 0.032 (0.010)     | 0.041 (0.009) |
| #2 | $\Delta_B$ | 0.037 (0.024)    | 0.095 (0.040)     | 0.108 (0.041) |
|    | $\Delta_C$ | 0.019 (0.016)    | 0.034 (0.009)     | 0.043 (0.009) |
| #3 | $\Delta_B$ | 0.030 (0.018)    | 0.092 (0.041)     | 0.098 (0.039) |
|    | $\Delta_C$ | 0.017 (0.016)    | 0.028 (0.011)     | 0.036 (0.010) |

## 6. Failure of Lu-Chipman decomposition of backscattering Mueller matrices

We provide simulated examples showing the failure of the Lu-Chipman decomposition of backscattering Mueller matrices. The first example is a backscattering Mueller matrix with a diattenuation vector $\boldsymbol{D} = (0.1, 0.1, 0.1)^T$, linear retardance $\delta=60°$, its orientation $\theta=-89°$, optical rotation $\Psi=6°$, and depolarization factors $d_0=1$, $d_1=1$, $d_2=-0.8$, and $d_3=-0.6$. Figure S6 and Table S5 show the results of Reciprocal polar decomposition and Lu-Chipman decomposition, respectively. The second example is identical to the first one, except the linear retardance $\delta$ is modified from 60° to 100°. Figure S7 and Table S6 show the corresponding results.

In both examples, the parameters recovered by reciprocal polar decomposition are in excellent agreement with the ground truth. In contrast, the Lu-Chipman decomposition yields erroneous values for the diattenuation vector, the linear retardance, the optical rotation, $d_2$, $d_3$, and depolarization anisotropy A. It also produces erroneous orientation $\theta$ for the retardance exceeding π/2. The behavior of the simulated data is consistent with the results reported in the main text for the birefringence resolution target, the tissue sample, and the chiral medium.

$\theta = -89°, \Psi = 6°, \delta = 60°$

$M_\Delta = diag(1,1,-0.8,-0.6)$

$M_R = \begin{pmatrix} 1 & 0 & 0 & 0 \\ 0 & 0.981 & 0.121 & -0.150 \\ 0 & -0.191 & 0.486 & -0.853 \\ 0 & -0.030 & 0.866 & 0.500 \end{pmatrix}$

$D = (0.1, 0.1, 0.1)^T$

$M_D = \begin{pmatrix} 1 & 0.100 & 0.100 & 0.100 \\ 0.100 & 0.990 & 0.005 & 0.005 \\ 0.100 & 0.005 & 0.990 & 0.005 \\ 0.100 & 0.005 & 0.005 & 0.990 \end{pmatrix}$

$M = M_D^\# M_R^\# M_\Delta M_R M_D$

$M_\Delta = \begin{pmatrix} 1 & 0 & 0 & 0 \\ 0.002 & 0.986 & 0.027 & 0.028 \\ 0.065 & 0.027 & \mathbf{0.647} & \mathbf{0.092} \\ -0.061 & 0.028 & \mathbf{0.092} & \mathbf{0.748} \end{pmatrix}$

$M_R = \begin{pmatrix} 1 & 0 & 0 & 0 \\ 0 & 0.998 & \mathbf{0.059} & \mathbf{0.035} \\ 0 & \mathbf{0.060} & \mathbf{0.500} & -0.864 \\ 0 & -0.033 & 0.864 & \mathbf{-0.503} \end{pmatrix}$

$M_D = \begin{pmatrix} 1 & \mathbf{0.203} & \mathbf{0.022} & \mathbf{0.084} \\ \mathbf{0.203} & 0.996 & 0.002 & 0.009 \\ \mathbf{0.022} & 0.002 & 0.976 & 0.001 \\ \mathbf{0.084} & 0.009 & 0.001 & 0.979 \end{pmatrix}$

$M = M_{mirror} M$

$M = M_\Delta M_R M_D$

$QM = [QM + (QM)^T]/2$

$QM = M_D^T M_R^T M'_{\Delta d} M_R M_D$

$M_\Delta = diag(1,1,-0.8,-0.6)$

$M_R = \begin{pmatrix} 1 & 0 & 0 & 0 \\ 0 & 0.981 & 0.121 & -0.150 \\ 0 & -0.191 & 0.486 & -0.853 \\ 0 & -0.030 & 0.866 & 0.500 \end{pmatrix}$

$M_D = \begin{pmatrix} 1 & 0.100 & 0.100 & 0.100 \\ 0.100 & 0.990 & 0.005 & 0.005 \\ 0.100 & 0.005 & 0.990 & 0.005 \\ 0.100 & 0.005 & 0.005 & 0.990 \end{pmatrix}$

(a)        (b)

Fig. S6. Decomposition of simulated backscattering Mueller matrices. (a) Lu-Chipman decomposition, and (b) Reciprocal polar decomposition. Erroneous values are highlighted in bold.

Table S5. The polarization parameters of decomposition of backscattering Mueller matrices. Erroneous values are highlighted in bold.

| Parameters | Backward Geometry | | Ground Truth |
|---|---|---|---|
| | Lu-Chipman | Reciprocal Polar | |
| **D** | **(0.20, 0.02, 0.08)$^T$** | (0.10, 0.10, 0.10)$^T$ | (0.10, 0.10, 0.10)$^T$ |
| $\theta$ (deg) | -88.9 | -89.0 | -89.0 |
| $\delta$ (deg) | **120.2** | 60.0 | 60.0 |
| $\Psi$ (deg) | **-0.1** | 6.0 | 6.0 |
| $\Delta$ | 0.21 | 0.20 | 0.20 |
| d1 | 0.99 | 1.00 | 1.00 |
| d2 | **0.65** | -0.80 | -0.80 |
| d3 | **0.75** | -0.60 | -0.60 |
| A | **0.21** | 0.11 | 0.11 |

$$M_\Delta = diag(1,1,-0.8,-0.6) \quad M_R = \begin{pmatrix} 1 & 0 & 0 & 0 \\ 0 & 0.985 & 0.004 & -0.171 \\ 0 & -0.168 & -0.177 & -0.970 \\ 0 & -0.034 & 0.984 & -0.174 \end{pmatrix} \quad M_D = \begin{pmatrix} 1 & 0.100 & 0.100 & 0.100 \\ 0.100 & 0.990 & 0.005 & 0.005 \\ 0.100 & 0.005 & 0.990 & 0.005 \\ 0.100 & 0.005 & 0.005 & 0.990 \end{pmatrix}$$

with $\theta=-89°, \Psi=6°, \delta=100°$ and $D=(0.1,0.1,0.1)^T$

$$M = M_D^\# M_R^\# M_\Delta M_R M_D$$

(a) Lu-Chipman decomposition:

$$M_\Delta = \begin{pmatrix} 1 & 0 & 0 & 0 \\ 0.007 & 0.989 & 0.006 & 0.039 \\ 0.058 & 0.006 & \mathbf{0.601} & -0.032 \\ -0.038 & 0.039 & -0.032 & \mathbf{0.794} \end{pmatrix}$$

$$M_R = \begin{pmatrix} 1 & 0 & 0 & 0 \\ 0 & 0.997 & \mathbf{0.073} & \mathbf{-0.011} \\ 0 & \mathbf{0.072} & \mathbf{-0.935} & \mathbf{0.347} \\ 0 & \mathbf{0.015} & \mathbf{-0.347} & \mathbf{-0.938} \end{pmatrix}$$

$$M_D = \begin{pmatrix} 1 & \mathbf{0.196} & \mathbf{0.073} & \mathbf{0.192} \\ \mathbf{0.196} & 0.973 & 0.007 & 0.019 \\ \mathbf{0.073} & 0.007 & 0.976 & 0.007 \\ \mathbf{0.192} & 0.019 & 0.007 & 0.979 \end{pmatrix}$$

With intermediate steps: $M = M_{mirror} M$, then $M = M_\Delta M_R M_D$

(b) Reciprocal polar decomposition:

$QM = [QM + (QM)^T]/2$, $QM = M_D^T M_R^T M'_{\Delta d} M_R M_D$

$$M_\Delta = diag(1,1,-0.8,-0.6)$$

$$M_R = \begin{pmatrix} 1 & 0 & 0 & 0 \\ 0 & 0.985 & 0.004 & -0.171 \\ 0 & -0.168 & -0.177 & -0.970 \\ 0 & -0.034 & 0.984 & -0.174 \end{pmatrix}$$

$$M_D = \begin{pmatrix} 1 & 0.100 & 0.100 & 0.100 \\ 0.100 & 0.990 & 0.005 & 0.005 \\ 0.100 & 0.005 & 0.990 & 0.005 \\ 0.100 & 0.005 & 0.005 & 0.990 \end{pmatrix}$$

Fig. S7. Decomposition of simulated backscattering Mueller matrices with an altered linear retardance. (a) Lu-Chipman decomposition, and (b) Reciprocal polar decomposition. Erroneous values are highlighted in bold.

Table S6. The polarization parameters of decomposition of backscattering Mueller matrices of an altered linear retardance. Erroneous values are highlighted in bold.

| Parameters | Backward Geometry | | Ground Truth |
|---|---|---|---|
| | Lu-Chipman | Reciprocal Polar | |
| **D** | **(0.20,0.07,0.19)$^T$** | (0.10,0.10,0.10)$^T$ | (0.10,0.10,0.10)$^T$ |
| $\theta$ (deg) | **1.2** | -89.0 | -89.0 |
| $\delta$ (deg) | **159.7** | 100.0 | 100.0 |
| $\Psi$ (deg) | **0.4** | 6.0 | 6.0 |
| $\Delta$ | 0.21 | 0.20 | 0.20 |
| d1 | 0.99 | 1.00 | 1.00 |
| d2 | **0.60** | -0.80 | -0.80 |
| d3 | **0.79** | -0.60 | -0.60 |
| A | **0.24** | 0.11 | 0.11 |

## 7. Failure of Symmetric decomposition of backscattering Mueller matrices

Symmetric decomposition [9] decomposes a Mueller matrix as the product

$$M = M_{D2} M_{R2} M_\Delta M_{R1}^T M_{D1} \tag{S26}$$

where $M_{D1}$ and $M_{D2}$ are the Mueller matrices of diattenuators, $M_{R2}$ and $M_{R1}^T$ are the Mueller matrices of retarders, and $M_\Delta = diag(d_0, d_1, d_2, d_3)$ is the Mueller matrix of a depolarizer. The diattenuator and retarder matrices, $M_D$ ($M_{D1}$ and $M_{D2}$) and $M_R$ ($M_{R2}$ and $M_{R1}^T$), take the same form as in Lu-Chipman decomposition. The polarization parameters can be calculated by using Eqs. (S4)-(S8). Using the symmetrical decomposition, we analyzed the measured backscattering Mueller matrix for the birefringence resolution target and beef tissue section. The two retarder matrices are expected to be close to each other and represent the retardance of the sample.

The extracted linear retardance and orientation angle, as well as depolarization for the target by the symmetric decomposition in the backward geometry, are shown in Fig. S8. Both the $M_{R1}$ and $M_{R2}$ matrices from the symmetric decomposition yield incorrect linear retardance and orientation angle. The extracted linear retardance, orientation angle, depolarization, and depolarization anisotropy for the fresh beef sections of thickness 100 μm and 300 μm by the symmetric decomposition in the backward geometry are shown in Fig. S9. Both the $M_{R1}$ and $M_{R2}$ matrices from the symmetric decomposition yield incorrect linear retardance and orientation angle. In addition, the image quality of polarization parameters obtained through symmetric decomposition is much poorer than those from reciprocal polar decomposition.

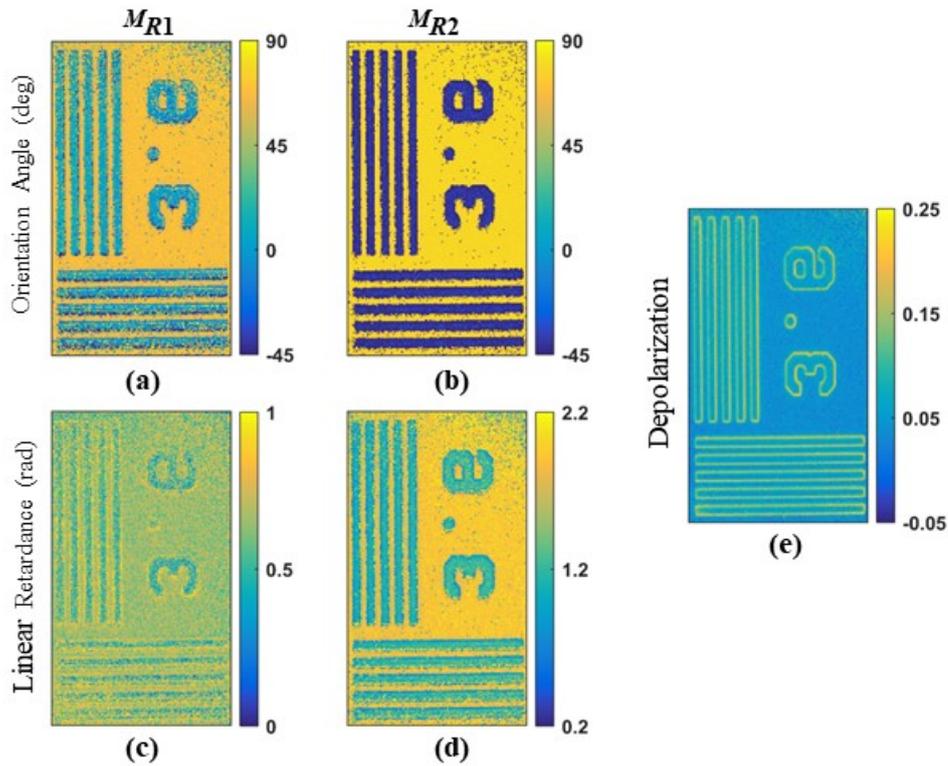

Fig. S8. The orientation angle (a, b), linear retardance (c, d), depolarization (e) by symmetric decomposition of the Mueller matrix for the target measured in the backward geometry. (a, c) for $M_{R1}$ and (b, d) for $M_{R2}$.

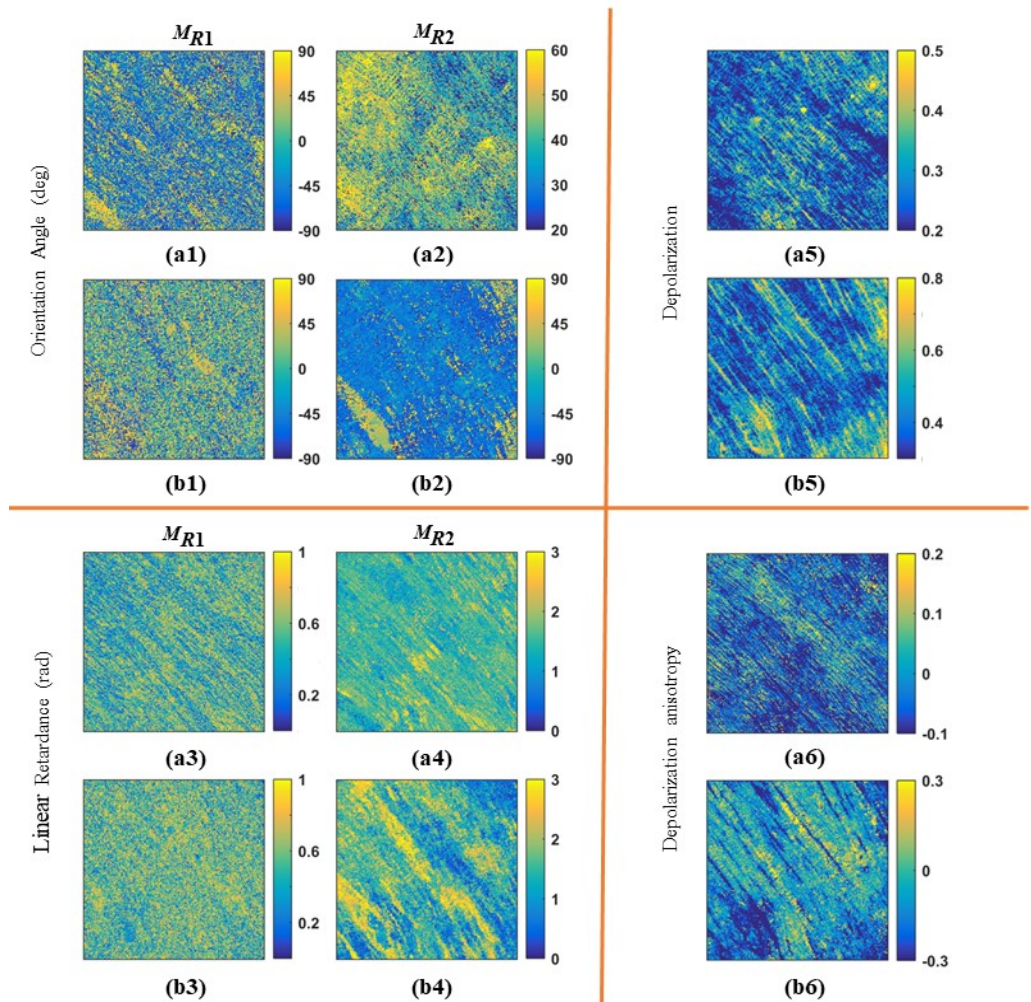

Fig. S9. The orientation angle, linear retardance, depolarization, and depolarization anisotropy by symmetric decomposition of (a1, a2, a3, a4, a5, a6) the 100-μm beef section and (b1, b2, b3, b4, b5, b6) the 300-μm beef section measured in the backward geometry. (a1, b1, a3, b3) for $M_{R1}$ and (a2, b2, a4, b4) for $M_{R2}$.